\documentclass[epj, final]{svjour}

\usepackage{palatino}
\usepackage[latin1]{inputenc}
\usepackage{epsf}
\usepackage{amsmath,amssymb}
\usepackage{latexsym}
\usepackage{calc}
\usepackage{color}
\usepackage{shadow}
\usepackage{epsfig}
\usepackage{float}

\usepackage[pdftex, pdfstartview = {FitH}]{hyperref}

\newcommand{\ben}{\begin{equation}}
\newcommand{\een}{\end{equation}}
\newcommand{\bea}{\begin{eqnarray}}
\newcommand{\eea}{\end{eqnarray}}

\def\sss{\scriptscriptstyle\rm}

\def\c{_{\sss C}}
\def\s{_{\sss S}}
\def\xc{_{\sss XC}}

\def\H{_{\sss H}}

\def\ext{_{\rm ext}}

\begin{document}
\title{Electron Scattering in Time-Dependent Density Functional Theory}

\author{Lionel Lacombe\inst{1} \and Yasumitsu Suzuki\inst{2} \and Kazuyuki Watanabe\inst{2} \and Neepa T. Maitra\inst{1,3}}
\institute{Department of Physics and Astronomy, Hunter College and the Graduate Center of the City University of New York, 695 Park Avenue, New York, New York 10065, USA
\and Department of Physics, Tokyo University of Science, 1-3 Kagurazaka, Shinjuku-ku, Tokyo 162-8601, Japan
\and The Physics Program and the Chemistry Program of the Graduate Center of the City University of New York, 695 Park Avenue, New York, New York 10065, USA} 
%\email{nmaitra@hunter.cuny.edu; 
\date{\today}
%\pacs{31.15.-p, 31.50.-x, 82.20.Gk}

\abstract
{ It was recently shown [Y. Suzuki, L. Lacombe, K. Watanabe, and N. T. Maitra, Phys. Rev. Lett. {\bf 119}, 263401 (2017)]  that peak and valley structures in the exact exchange-correlation potential of time-dependent density functional
 theory are crucial for accurately capturing time-resolved dynamics of electron scattering in a model one-dimensional system. Approximate functionals used today miss these structures and consequently underestimate the scattering probability. The dynamics can vary significantly depending on the choice of the initial Kohn-Sham state, and, with a judicious choice, a recently-proposed non-adiabatic approximation provides extremely accurate dynamics on approach to the target but this ultimately also fails to capture reflection accurately. Here we provide more details, using a model of electron-He$^+$ as illustration, in both the inelastic and elastic regimes. In the elastic case, the time-resolved picture is contrasted with the time-independent picture of scattering, where the linear response theory of TDDFT can be used to extract transmission and reflection coefficients. Although the exact functional yields identical scattering probabilities when used in this way as it does in the time-resolved picture, we show that the currently-available approximate functionals do not, even when they have the correct asymptotic behavior. }

\maketitle
\section{Introduction}

Electron scattering is, in a sense, the godmother of time-dependent density functional theory (TDDFT). In the 1980's, Hardy Gross, a postdoctoral fellow at the Institute for Theoretical Physics at the Goethe
University Frankfurt, where nuclear physics and atomic scattering were  major themes of research, pondered the question: When an electron scatters from an ion, what is the time-dependent potential that drives its motion? The theorems of Hohenberg, Kohn, and Sham, had, twenty years earlier, presented a potential of this spirit for an electron in a ground-state, but can such a potential be defined when the electron is undergoing the intricate dance with all the other electrons and the nucleus in a scattering event?  This question led eventually to the birth of the Runge-Gross theorem~\cite{tddft1}, where Hardy, and student Erich Runge, answered the question affirmatively. Their theorem states that all properties of an interacting many-body system evolving from a given initial state, can be found from knowledge of the one-body density alone. One can then obtain all properties of interest from a non-interacting system that reproduces this density, and the potential in that system, called the Kohn-Sham potential, is the one at the root of Hardy's questions many years ago. Since then, TDDFT has grown into a successful and well-established method for electronic excitations and dynamics~\cite{tddft1,tddft2,tddft3},  enabling calculations on systems impossible to study otherwise. 

Not surprisingly, this exact potential is difficult to find for electron scattering off realistic targets in the most interesting situations.  Electron scattering is ubiquitous in physics, chemistry, and biology, in both nature (e.g. Ref.~\cite{Sanche}) as well as in experimental techniques that probe matter (e.g. Ref.~\cite{Crommie}), so theoretical methods that can accurately describe the process without having to solve the computationally expensive highly-correlated, non-perturbative, many-body problem directly, are of great interest. TDDFT, with approximations to the Kohn-Sham potential, has been applied to real-time non-perturbative calculations of protons and anti-protons scattering from small molecules~\cite{GWWZ14,QSAC17,Kirchner} and of electron wavepacket scattering from graphene~\cite{scat3,scat4,scat5,scat6}, and been applied to compute elastic electron-atom scattering cross-sections
by means of linear-response theory~\cite{FWEZ07,FB09,WMB05}. Increasingly, a time-resolved picture is needed. The agreement with experimental results is often good, but not always, and an understanding of what aspect of the approximate potential is causing the error is desirable.
Recently, an examination of how well the approximations to the time-dependent Kohn-Sham potential were working revealed large discrepancies from the exact potential in some two-electron one-dimensional models~\cite{SLWM17}. It was found that the TDDFT approximations in use today are missing peak and valley features that largely influence the scattering process. Ref.~\cite{SLWM17} showed that to capture these features, one must go beyond the usual adiabatic approximations of TDDFT, but that even a recently proposed non-adiabatic functional misses them. The adiabatic approximations show quite unusual dynamics, including spurious density oscillations, depending on how the Kohn-Sham initial state is chosen. In this paper, we flesh out some of the details of the calculations but illustrated instead on e-He$^+$ scattering. The model is presented in Sec.~\ref{sec:model}, followed by the exact and approximate TDDFT descriptions in Sec.~\ref{sec:tddft}. Sec.~\ref{sec:lrtddft} shows how to obtain transmission and reflection probabilities from linear response theory in the elastic scattering regime and compares the results with the fully time-resolved calculations. We provide a summary and some conclusions in Sec.~\ref{sec:outlook}.

\section{1D Electron-Ion Scattering Model}
\label{sec:model}
We consider a model of two-electron scattering in one dimension. 
One of the electrons begins in the ground-state $\phi_{\rm gs}(x)$ of an external potential 
 $v_{\rm ext}(x)=-\frac{Z}{\sqrt{(x+10)^2+1}}$ that is 
 a soft-Coulomb model of either
a Hydrogen atom with $Z=1$ (previous work \cite{SLWM17}) or a singly-ionized Helium atom with $Z=2$. We focus on the latter in this paper. 
We use atomic units in this paper unless otherwise stated.
The ion is localized at $x=-10$~a.u. and is the target of 
the other electron, which begins in a Gaussian wavepacket localized at $x=10$ with a certain velocity $p$:
\ben
\begin{split}
\phi_{\rm WP}(x)=\left(2\alpha/\pi\right)^{\frac{1}{4}}e^{\left[-\alpha(x-x_0)^2+ip(x-x_0)\right]}\,.
\end{split}\label{eqn: wp}
\een
We take $\alpha = 0.1$ and take $p=0.6$ to study 
elastic scattering and $p=1.2$ for inelastic scattering. The lowest singlet excitation of the He$^+$ target is $0.71$ a.u., which would correspond to an incoming moment of $1.19$ a.u. to excite. 
The two electrons are subject to a soft-Coulomb repulsive interaction $W_{ee}(x_1,x_2)=\frac{1}{\sqrt{(x_1-x_2)^2+1}}$.
Thus the Hamiltonian of this system is the following: 
\ben
{\hat H}(x_1,x_2)=\sum_{i=1,2}\left(-\frac{1}{2}\frac{\partial^2}{\partial x_i^2}+v_{\rm ext}(x_i)   \right) +  W_{ee}(x_1,x_2)
\een
and the initial spatial part of the interacting wavefunction is
\ben
\Psi_0(x_1,x_2) = \frac{1}{ \sqrt{2}}\left(\phi_{\rm gs}(x_1)\phi_{\rm WP}(x_2) +\phi_{\rm WP}(x_1)\phi_{\rm gs}(x_2) \right)
\label{eqn:Psi}
\een
where we chose a singlet state for the spin part as indicated by the $+$ sign and
 $\phi_{\rm gs}$ is the ground-state of one electron alone in 
the external potential $v_{\rm ext}$.
This state is propagated by numerically solving the full 
time-dependent Schr\"odinger equation $i\partial_t\Psi(x_1,x_2,t)={\hat H}(x_1,x_2)\Psi(x_1,x_2,t)$ which is tractable in 1D.

\begin{figure}[h]
 \centering
\includegraphics*[width=1.0\columnwidth]{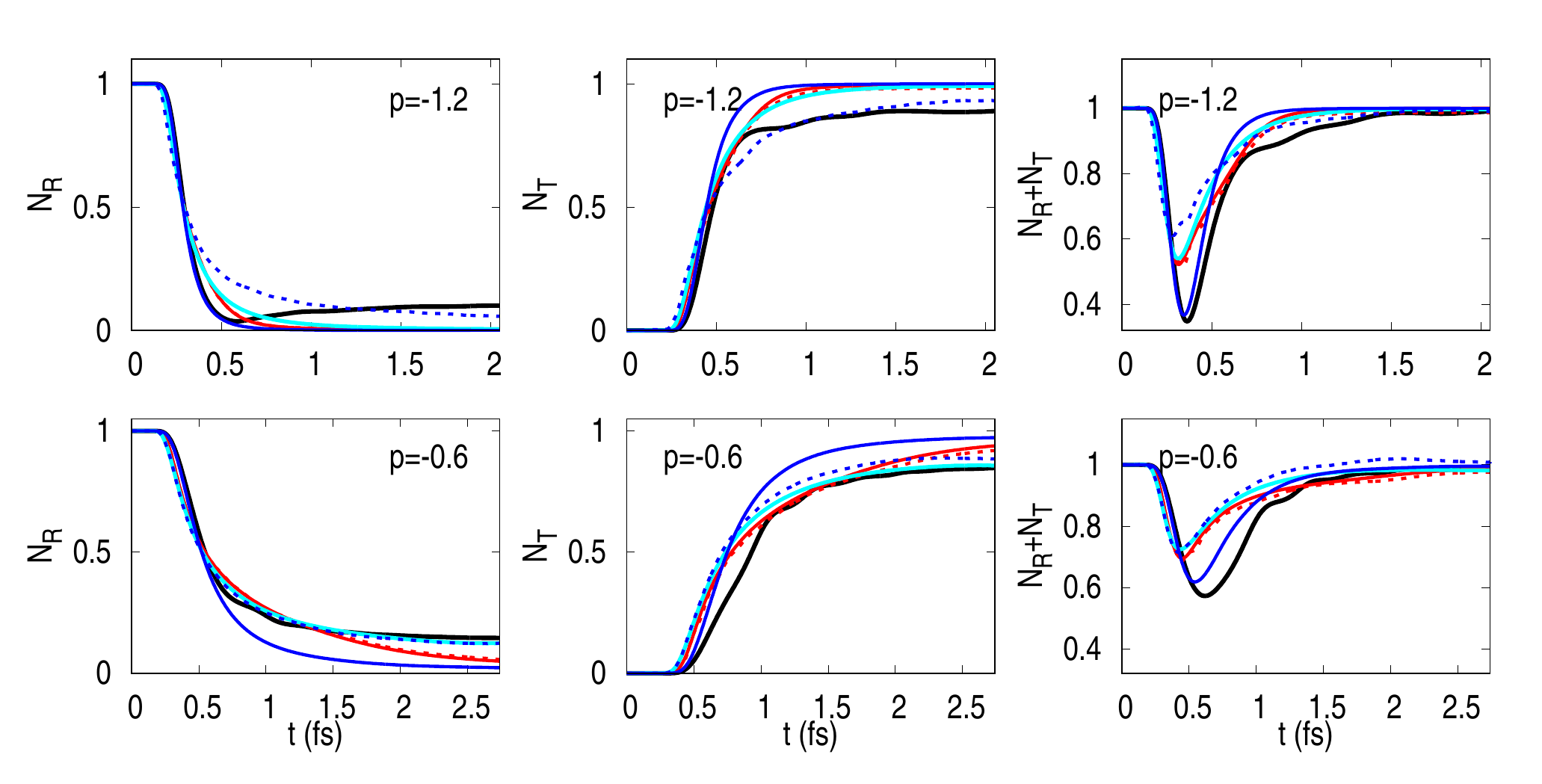}
\caption{(color online). Number of  electrons in the reflection region $N_R$ (left panel) and transmission region $N_T$ (middle panel) and $N_R + N_T$ (right panel) for the exact
 (black solid), ALDA ($\Phi_0^{(1)}$: red dashed, $\Phi_0^{(2)}$: red solid), 
 $v_{\rm xc}^{\rm S}$ ($\Phi_0^{(1)}$: blue dashed, $\Phi_0^{(2)}$: blue solid),
 AEXX ($\Phi_0^{(1)}$: equal to $v_{\rm xc}^{\rm S}$ (blue dashed), $\Phi_0^{(2)}$: cyan solid)
 for the two different momenta
$p=-1.2$ (upper panels) and $p=-0.6$ (lower panels). 
} 
  \label{fig:NRNT}
\end{figure}

We study the electronic density $n(x,t)=2\int |\Psi(x,x_2,t)|^2dx_2$
and the  number of electrons transmitted(reflected), 
$N_T(N_R)$ which is defined as the integral of 
the density over the region $x>-5.0$ ($x<-15$). 
In all the calculations we used a box of $200$a.u. with reflecting boundaries but 
absorbing boundaries based on a mask method give the same results over the time period shown. 
The black curves in Fig.~\ref{fig:NRNT} show $N_R(N_T)$ and their sum for $p=1.2$ and $p=0.6$.
Similar plots for e-H, but for different incoming momenta, can be found in Fig 2 of Ref.~\cite{SLWM17}; generally, for a fixed incoming momentum, there is more scattering off the H atom than there is for the He$^+$ atom.
%likely because the stronger ionic potential in the latter cases confines the target density more closely.
$N_R$ starts at one as initially the incoming electron is localized on the right side ($x>-5$)
of the box. 
Then $N_R$ approaches zero as the electron enters the target region ($-15<x<-5$) and 
increases again as part of the density is reflected until it reaches an asymptotic limit.
Similarly $N_T$ starts at zero and increases as the electron is transmitted.
We see for the case of the higher incoming momentum, a fraction of charge gets trapped for some time in the ion before eventually leaving the ion with one electron as it had initially (although energized, as we will see from density oscillations in the target at long times). %For lower momenta (but still above the lowest target excitation), a fraction of the incoming charge does remain trapped in the ion at long times {\color{magenta} {\it check! i remember something like this in an e-discussion we had {\tiny(probability amplitude...)}}}.
% {\color{cyan} 1.19 is the lowest above excitation and if you look at NR+NT you just see that in the exact case it takes very long time to really go to 1 but it is a small effect. I think we should talk about this as I am not even sure the calculations are accurate enough to interpret such a small event}
In the e-H case of Ref.~\cite{SLWM17} on the other hand, the scattering interaction ends up slightly ionizing the target. 
In both cases, the scattering is instead elastic for the lower momenta shown. The other curves shown in this plot correspond to TDDFT approximations which we will discuss in the next section. 

\begin{figure}[h]
 \centering
%\boxed{
 \includegraphics*[width=1.0\columnwidth]{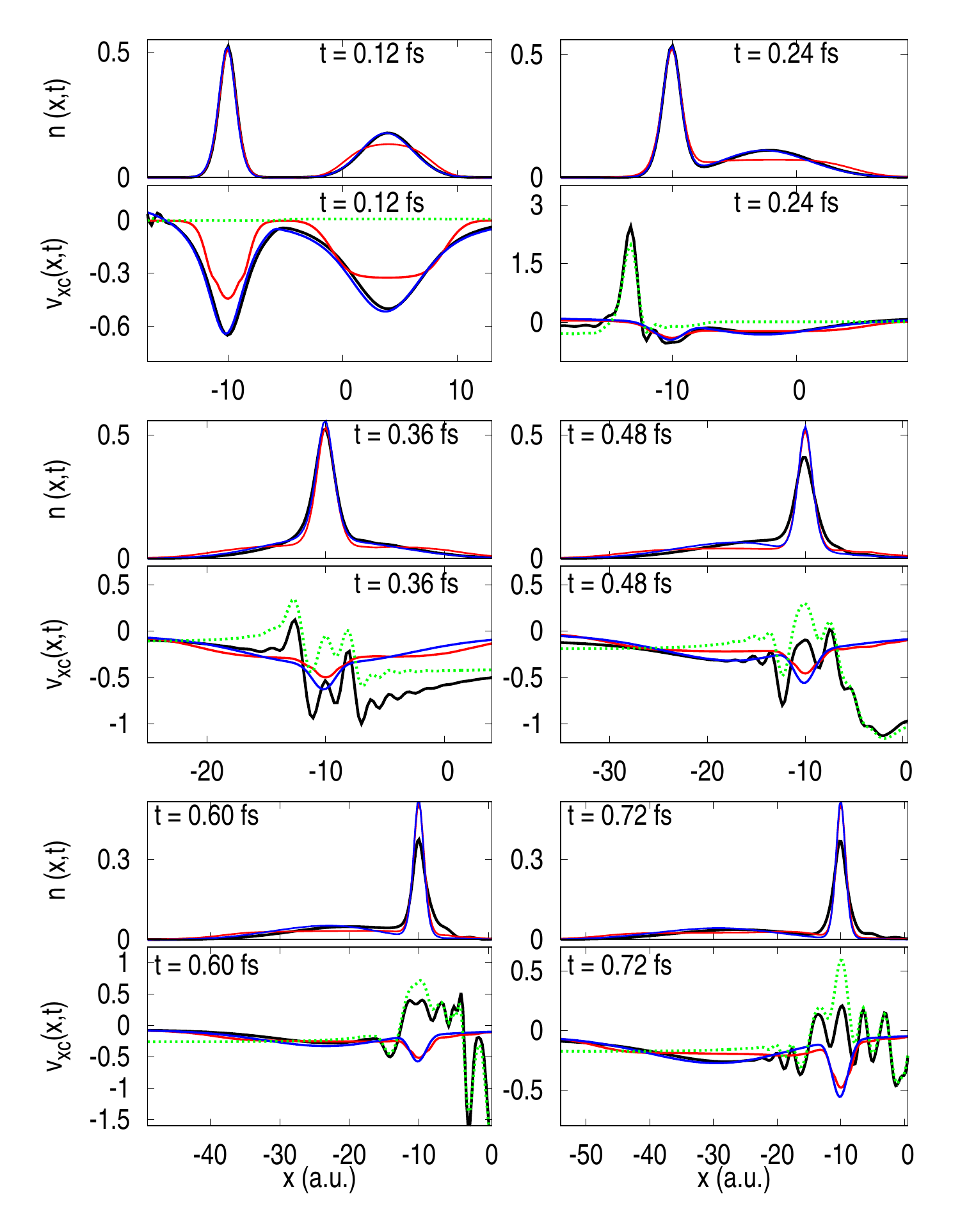}
 %}
 \caption{(color online). Snapshots of the exact electron density $n (x,t)$ in the e-He$^+$ inelastic scattering model system ($p = 1.2$au)
 (black solid line in the upper panel for each time slice).
 Black line in the lower panel shows the exact time-dependent xc potential $v_{\rm xc}$ for the initial KS state $\Phi_0^{(1)}$ for each time slice.
The results of ALDA (red solid line) and $v^{\rm S}_{\rm xc}$ (blue solid line) are shown in each panel.
The kinetic component of the exact xc potential $v^{\rm T}_{\rm c}$ is also shown as green dotted line in the lower panels.}
 \label{fig:inelastic_phi1}
\end{figure}

\begin{figure}[h]
 \centering
 \includegraphics*[width=1.0\columnwidth]{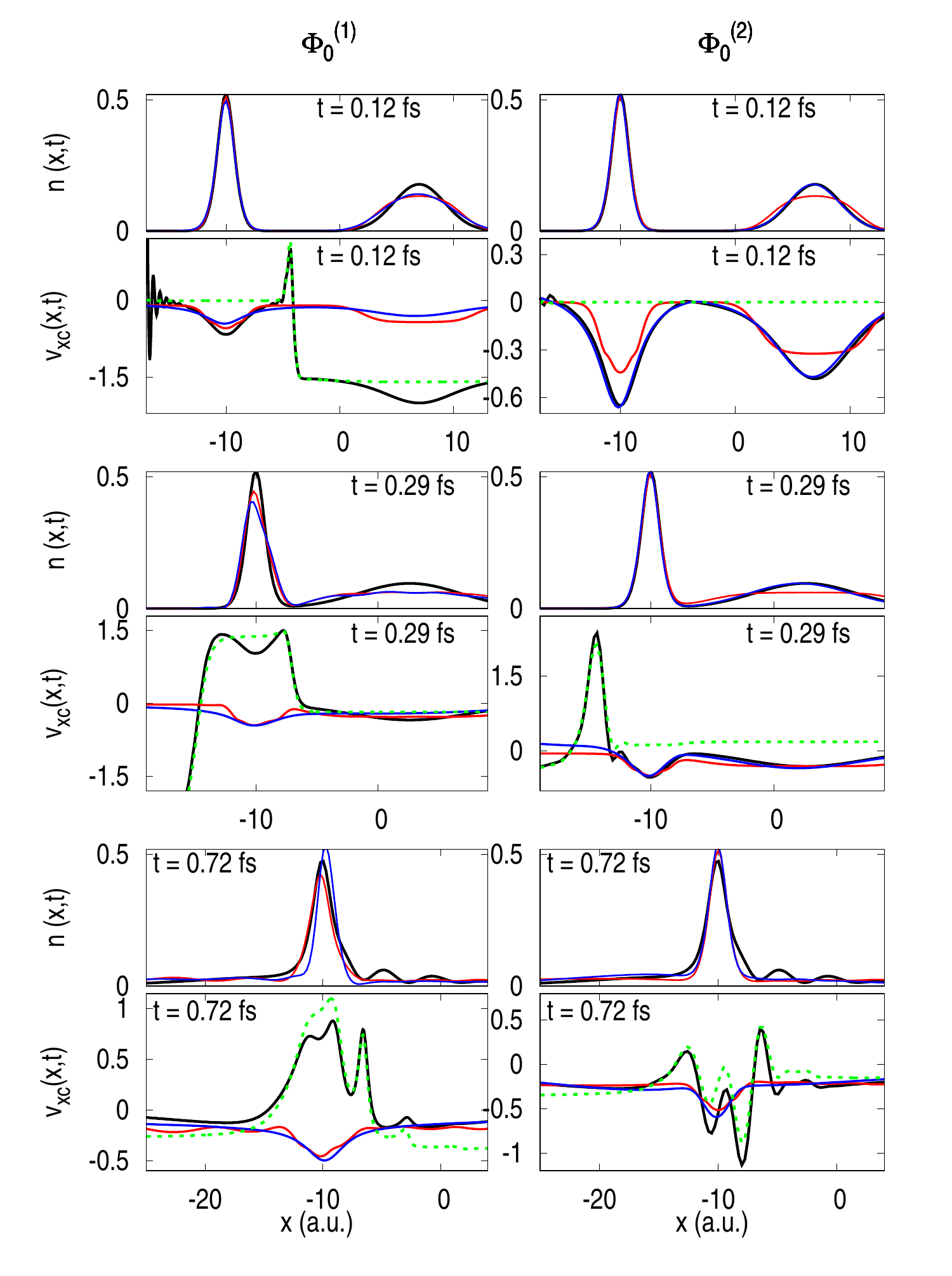}
 \caption{(color online). Snapshots of the exact electron density $n (x,t)$ in the e-He$^+$ elastic scattering model system ($p = 0.6$au)
 (black solid line in the upper panel for each time slice).
Black line in the lower panel shows the exact time-dependent xc potential $v_{\rm xc}$ for the initial KS state $\Phi_0^{(1)}$ (left column) and $\Phi_0^{(2)}$ (right column) for each time slice.
The results of ALDA (red solid line) and $v^{\rm S}_{\rm xc}$ (blue solid line) are shown in each panel.
The kinetic component of the exact xc potential $v^{\rm T}_{\rm c}$ is also shown as green dotted line in the lower panels. }
 \label{fig:elastic}
\end{figure}

The exact density is plotted as the black solid line at different time slices in the upper panels of 
Fig.~\ref{fig:inelastic_phi1} (and Fig.~\ref{fig:inelastic_phi2}) for $p=1.2$~a.u. and Fig.~\ref{fig:elastic} for $p=0.6$~a.u.
In the case of $p=1.2$~a.u., after the collision the density remaining in the target is more spread than it was initially, as the target was left
excited by the collision, confirming we are in an inelastic scattering situation; in fact the density in the target is in a non-stationary state. Similar density oscillations were seen at long times for inelastic e-H scattering studied in Ref.~\cite{SLWM17}. 
%{\color{magenta} A question is whether we make the movies for this case too?}
On the other hand, in the elastic $p=0.6$~a.u. case of Fig.~\ref{fig:elastic},
the final target density is the same as the initial one of ground-state wavefunction, meaning
 no energy has been transmitted to the target.

Shown also in these figures are the results from TDDFT approximations (red and blue lines), 
neither of which yield reflection nor the dynamics correctly.
In Ref.~\cite{SLWM17} we showed why using the case of e-H scattering: the exact xc potential contains crucial peak and valley structures that are largely responsible for scattering but that are missing in the approximations. 

\section{TDDFT: exact and approximate}
\label{sec:tddft}
As in DFT, the principle of TDDFT is to map 
an interacting system on a non-interacting one with an effective potential, $v\s$, in which the non-interacting fermions evolve with the same one-body density $n(x,t)$ as that of the interacting system. 
The non-interacting system is propagated under
the single-particle Hamiltonian, 
\ben
\hat{h}^s(x_1,x_2) = 
\sum_{i=1,2} \left( -\frac{1}{2}\frac{\partial^2}{\partial x_i^2} + v\s(x_i,t) \right)\;,
\een
written for two electrons in 1D, where 
\ben
v_s[\Phi_0,n] = v_{ext}[\Psi_0,n] + v_H[n] + v_{xc}[n;\Psi_0,\Phi_0]\;.
\een
Here $v\H$ is the Hartree potential and $v\xc$ is the exchange-correlation (xc)
potential. The indicated functional dependences follow from the Runge-Gross one-to-one density-potential mapping that holds for a given initial state~\cite{tddft1,tddft3}.
In particular, it is important to note the dependence of the xc potential on the
initial interacting wavefunction $\Psi(0)=\Psi_0$ and KS wavefunction $\Phi(0)=\Phi_0$; the exact xc potential can be very different for different initial states that share the same one-body density.  

Numerically, if one has the exact evolution of the density arising from a fixed initial state $\Psi_0$
(by propagating the exact solution), it is possible to compute
the exact time-dependent $v\s(x,t)$ for a given $\Phi_0$ using the global fixed-point iteration method of Ref.~\cite{NRL13,RPL15}. 
A sketch of this method is as follows:
at each time-step the Kohn-Sham system is propagated with an initial guess for the potential. Then
its density is compared with the reference density of the exact system and 
the potential is then modified according to Eq. (10) of Ref.~\cite{NRL13}.
These steps are repeated until convergence of the density for this time-step. 
In our case we limited the number of iterations as this algorithm 
has the tendency of over-fitting the noise at low density; a check can always be performed via checking the density after propagating the system with the potential that is found in this way. 
Error builds up over time between the simulated and reference densities, which was the main issue that limited the simulation time we could reach. The results presented here are obtained before the KS density and the exact start to diverge.

\begin{figure}[h]
 \centering
 \includegraphics*[width=1.0\columnwidth]{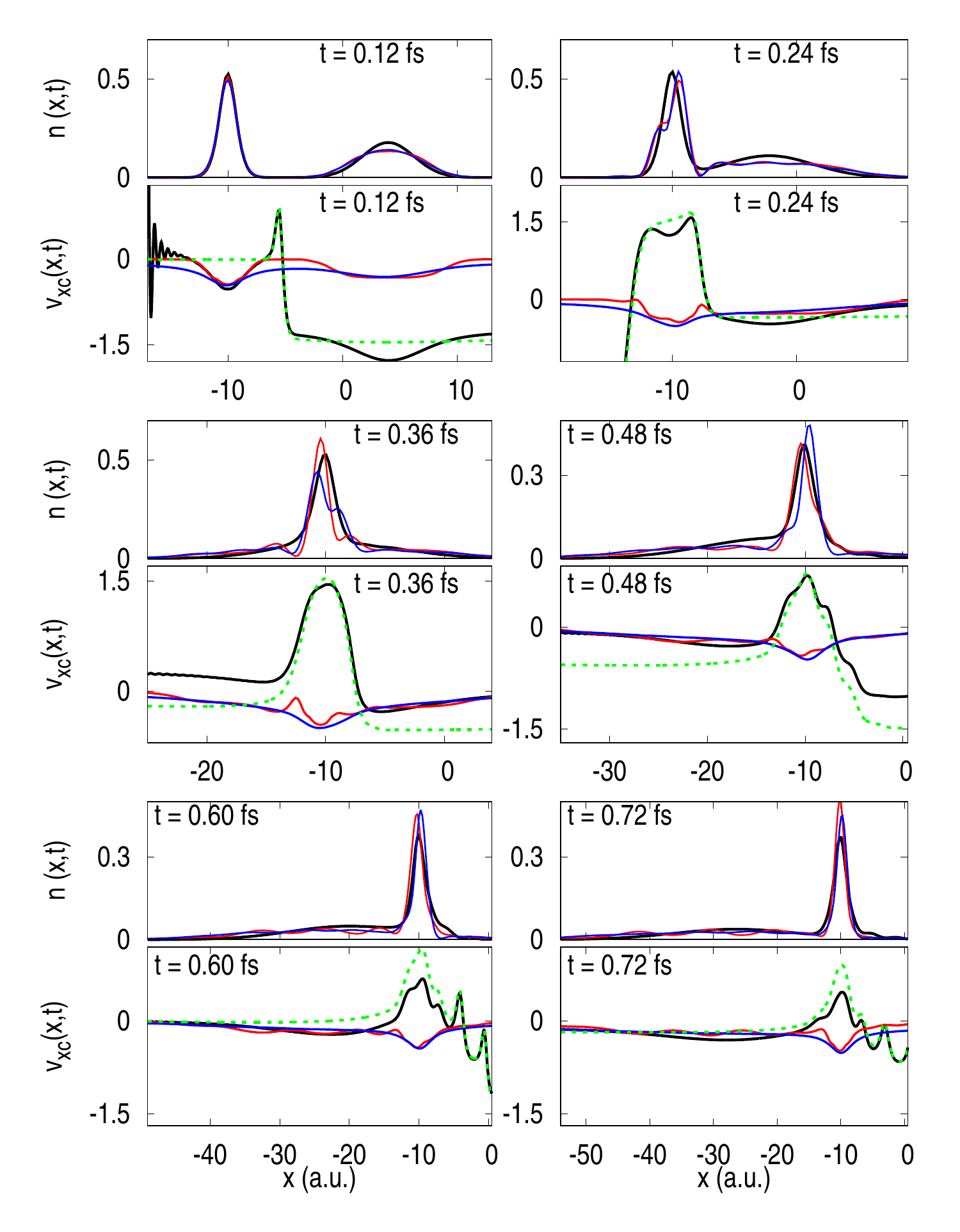}
 \caption{Same as Fig.~\ref{fig:inelastic_phi1}, but for the initial KS state $\Phi_0^{(2)}$. }
 \label{fig:inelastic_phi2}
\end{figure}

Using this method, we compute $v\xc(x,t)$ which is plotted in black in the lower panels of Figs.~\ref{fig:inelastic_phi1},~\ref{fig:elastic},~\ref{fig:inelastic_phi2}. 
In many time-dependent calculations the $\Psi_0$  is a ground-state wavefunction and $\Phi_0$ is naturally chosen also as a non-interacting
ground-state wavefunction; in this case, the initial states are themselves functionals of the density only.
In our case of time-resolved scattering, the initial physical state, $\Psi_0$, describes a wavepacket approaching a target, which is far from a ground-state, and initial-state dependence plays a paramount role. 
For $\Phi_0$ one can pick any state that has the same density $n_0(x)$ and 
the same first time-derivative of the density $\partial_t n_0(x)$ as
$\Psi_0$.

We consider two possibilities for 
the spatial part.
One is a single determinant with a doubly-occupied orbital: 
\ben
\Phi_0^{(1)}(x_1,x_2)
=\phi_0(x_1)\phi_0(x_2) 
\label{eqn: Phi_1}
\een
where imposing the restrictions on the density and its time-derivative leads to 
$\phi_0(x)=\sqrt{\frac{n_0(x)}{2}}\exp\left[i\int^x \frac{j_0(x')}{n_0(x')}dx'\right]$, with
$n_0(x)$ and $j_0(x)$ the initial density and current density of the interacting system. 
For the doubly-occupied state an analytical 
formula for $v_{xc}$ is straightforward to derive~\cite{xc1}.
It is a valid KS state, and, being a Slater determinant, perhaps a natural choice for a non-interacting system, despite having a structure very far from that of the interacting wavefunction.

The other choice we consider is simply the exact interacting state:
\ben
\begin{split}
\Phi_0^{(2)}(x_1, x_2)=\Psi_0&(x_1, x_2)\;,
\end{split}\label{eqn: Phi_11}
\een
which, although not a Slater determinant, obviously fulfills the required conditions.

The black solid lines in the lower panels of Figs.~\ref{fig:inelastic_phi1} and~\ref{fig:inelastic_phi2} show the snapshots of the exact xc potentials
$v\xc[n;\Psi_0,\Phi_0^{(1)}](x,t)$ and $v\xc[n;\Psi_0,\Phi_0^{(2)}](x,t)$ respectively, for the e-He$^+$ model. As observed in the case of e-H scattering in Ref.~\cite{SLWM17}, we observe peak and valley structures from when the electron approaches its target onwards. These play a crucial role in capturing the reflection accurately. 

For the case of the single-determinant, $\Phi_0^{(1)}$, (Fig.~\ref{fig:inelastic_phi1}), $v_{\rm xc}[n;\Psi_0,\Phi_0^{(1)}] (x,t)$
develops dynamical peak and step structures throughout the dynamics, even at very early times. One could argue that at very early times, at the first time-slice shown and earlier, the step and peak appear in a region of very little density, in between the incoming electron's wavepacket and the target, and so do not impact the ensuing dynamics very much. In fact the over-spreading of the incoming electron density produced from the approximate methods (more shortly) that is observed in the first time-slice shown is a fault of local properties of the approximations, rather than the lack of the step and the peak. However as soon as there is some overlap between the target and incoming densities, these structures do appear in regions where there is appreciable density, and they have an important  influence on the dynamics, in particular acting as dynamical barriers that reflect electron density back in the incoming direction, and also preventing unphysical interferences. As we shall see, approximations that miss these structures vastly underestimate the scattering probability. 
The exact xc potential for $\Phi_0^{(2)}$ (Fig.~\ref{fig:inelastic_phi2}) has no structure
 at very early times but displays a large peak structure behind the center of the
 target during the approach (at around $t=0.24$ fs for $p=1.2$~a.u. and $t=0.29$ fs for $p=0.6$~a.u.), and complicated peak and valley structures after the electron reaches the interaction region.
These structures were also observed
in the e-H scattering of Ref.~\cite{SLWM17}.  They appear here for both inelastic 
($p=1.2$~a.u.) and  elastic ($p=0.6$~a.u.) scattering.
The latter case is plotted in Fig.~\ref{fig:elastic} for both wavefunctions. 

\subsubsection{Approximate functionals}
\label{sec:approx}
We now consider propagation under two approximations commonly used in TDDFT.
The first one is the adiabatic local density approximation (ALDA), which is 
developed from the one-dimensional uniform gas~\cite{CSS06,1dldax}.
The ALDA results are plotted in red in Figures~\ref{fig:NRNT}--\ref{fig:inelastic_phi2} of this paper. 

As shown by Figs.~\ref{fig:inelastic_phi1}--\ref{fig:inelastic_phi2}, the ALDA potential causes the density to spread even before 
it interacts with the target, as mentioned above. Although one does expect some diffusion of a Gaussian wavepacket far from any target, under ALDA this is grossly exaggerated, and is due to the self-interaction error of ALDA which makes it propagate any locally one-electron system poorly. Further, when propagating the Slater determinant state $\Phi_0^{(1)}$, the ALDA density develops oscillations in time. This is due to unphysical interferences  of the electron with itself: the Kohn-Sham wavefunction describes each electron as delocalized on the target {\it and} on the incoming wavepacket. 
In fact, such a description is in itself perfectly allowed in exact TDDFT, for which the step and the peak in the {\it exact} xc potential between the two parts of the density seen at early times compensate the interference effect. Lacking this feature, the ALDA density exhibits spurious oscillations and the behavior
of the potential follows closely the local behavior of the density.
Contrary to the exact potential, ALDA does not display the complex features 
of the exact potential before or during the collision and is much smoother. ALDA propagation of $\Phi_0^{(2)}$ does not result in density oscillations as this state allows for a separate orbital for the target electron and the incoming one, but still shows the over-spreading due to self-interaction error. Finally, without the peak and valley structures, ALDA fails to capture much reflection. 

This is evident from Fig.~\ref{fig:NRNT}, where in the inelastic case, $N_R$ predicted from ALDA collapses to zero while $N_T$ approaches 1. In the elastic case, although not zero, ALDA significantly underestimates the reflection.  Choosing $\Phi^{(2)}$ (red, solid) over $\Phi^{(1)}$ (red, dashed)   does not have much effect on the integrated quantities $N_R$ and $N_T$ in this case. 

 We next consider a relatively new approximation arising from an exact decomposition of the xc potential into kinetic and interaction terms~\cite{kine1,kine2,xc2,xc3}. This is  $v\xc^{\rm S}$~\cite{xc3,SLWM17}, defined as
\ben
\begin{split}
v\xc^{\rm S}(x,t)=\int^x dx''\int n\xc^{\rm S}(x',x'',t)\frac{\partial}{\partial x''}W_{ee}(|x'-x''|)dx'\;,
\end{split}\label{eqn: vxcS}
\een
where $n\xc^{\rm S}(x',x'',t)$ is the xc hole of the Kohn-Sham system. This is an approximation to the interaction term of the exact xc potential, which completely neglects the kinetic term, $v\c^T$. 
For a Slater-determinant choice of Kohn-Sham wavefunction, $v\xc^{\rm S}$ reduces to time-dependent exact exchange (TD EXX), which, in turn, for two electrons, is equivalent to adiabatic exact-exchange (AEXX). So propagating $\Phi_0^{(1)}$ with $v\xc^{\rm S}$ is identical to AEXX. For a more general initial Kohn-Sham wavefunction (including $\Phi^{(2)}$),  $v\xc^{\rm S}$  includes some correlation, and yields an orbital-dependent functional, that
generally has spatial- and time-nonlocal dependence on the
density. This is because the xc hole depends instantaneously on the orbitals,
each of which has a time-nonlocal dependence on the
density and on the KS initial state. 

Figure~\ref{fig:inelastic_phi1} shows that propagation of $\Phi^{(1)}$ under $v\xc^{\rm S}$ (i.e. AEXX) is similar to the propagation using ALDA, in that the density over-spreads initially, develops oscillations, and the time-resolved $N_R$ and $N_T$ do not track the exact ones well for the inelastic case. Although at the later times shown in Fig.~\ref{fig:NRNT}, there is some probability of finding the electron in the reflected region, it actually continues to decay to 0 over long times, unlike the exact. 

For the elastic scattering case, AEXX appears to do very well when propagating the doubly-occupied orbital, but its success is not robust and might be somewhat misleading for a few reasons. First, this is a lucky accident since in fact AEXX does {\it not} typically capture time-resolved scattering well, as was clear in Ref.~\cite{SLWM17}, where it severely underestimated e-H reflection probabilities in both inelastic and elastic cases; see also Sec.~\ref{sec:lrtddft} and Fig.~\ref{fig:eH_aexx_exact}. Second, the AEXX is quite sensitive to the definition of the target region: if instead of integrating from $x=-5$ to the right-hand boundary, we integrate from $x=5$,  at large times the exact $N_R$ still settles to at about 0.16, while with  AEXX, $N_R$ drops to about 0.1. Similarly, if instead of integrating from $x = -15$ to the left-hand boundary, we integrate from $x = -25$, we find that asymptotically the exact $N_T$ remains at about 0.84 while the AEXX $N_T$ drops from about $0.86$ to $0.76$. Further, carrying out the propagation for longer times, the $N_R$ and $N_T$ of AEXX oscillates with an amplitude of about 0.05. A closer look at the ALDA and exact densities shows they are very different; not only is the AEXX more diffuse throughout the dynamics, there are distinct scattering wavepackets visible in the exact calculation but much more indistinct in the AEXX. 

For the initial Kohn-Sham state $\Phi_0^{(2)}$ the $v\xc^{\rm S}$ potential is actually exact at early times (c.f. blue curves on Fig~\ref{fig:inelastic_phi2}) and only starts diverging noticeably from the exact potential significantly into the collision. 
Because of the integrated structure of equation (\ref{eqn: vxcS}), 
this approximation stays very smooth. 
The main complexities of the exact xc potential are in fact contained in the kinetic component
$v_{c}^{T}$, missing in this approximation; this is shown as the green dashed curve on Fig~\ref{fig:inelastic_phi1},\ref{fig:elastic},\ref{fig:inelastic_phi2}.
Similarly to ALDA and AEXX, this approximation cannot reproduce the 
transmission and reflection coefficients of the exact potential despite $v\xc^{\rm S}$ and AEXX having 
the correct asymptotic behavior that ALDA lacks. 

All the observations in this section for e-He$^+$ scattering, regarding the importance of the choice of the Kohn-Sham initial state and the peak and valley structures seen in the exact xc potential that are required to accurately capture the reflection, echo  those for the e-H scattering that was found in Ref.~\cite{SLWM17}. These peak and valley structures are non-adiabatic features of the exact potential; they are missing in the "best" adiabatic approximation, as was shown in Ref.~\cite{SLWM17}, which is when the exact ground-state xc potential is used in the time-propagation. Although the relatively new approximation, $v\xc^{\rm S}$ is non-adiabatic, it does not improve the description of scattering, although it does improve the initial approach of the electron to the atom. Ref.~\cite{SLWM17} identified that a good approximation to the kinetic part of the exact xc potential, $v\c^T$, is required to accurately capture scattering dynamics in TDDFT. An expression is known for this in terms of the difference between the interacting and Kohn-Sham one-body density-matrices, so this needs somehow to be density-functionalized. 

One can also consider triplet scattering within the simple two-electron models. 
In this case, the spatial part of the initial interacting wavefunction $\Psi_0(x_1,x_2)$ reads
\ben
   \Psi_0(x_1,x_2) =
   \frac{1}{ \sqrt{2}}\left(\phi_{\rm gs}(x_1)\phi_{\rm WP}(x_2)
   -\phi_{\rm WP}(x_1)\phi_{\rm gs}(x_2) \right)
\een
and it is natural to consider the initial Kohn-Sham wavefunction $\Phi_0^{(2)}(x_1,x_2) = \Psi_0(x_1,x_2)$.
(In the triplet scattering case, because of the antisymmetry in the spatial part,
the doubly-occupied state $\Phi_0^{(1)}$ is forbidden). 
Only the sign has been changed and this sign has no effect on the initial
density nor on the Kohn-Sham single-particle orbitals that compose $\Phi_0^{(2)}$. 
This means that in fact this sign is invisible for spin-unpolarized 
ALDA propagation which gives the same results as the singlet case.
On the other hand, the exact results and $v\xc^{\rm S}$ both give
different results in the
triplet case due to their dependence on the wavefunction.  In the e-H triplet 
  scattering we find again the peak and valley structures appear in the exact potential, although they are smaller than those in the singlet case and the resulting reflection is also less than the singlet case, for a given incoming momentum. 
  %The dynamics produced by $v\xc^S$ is quite different from the exact but both lead to an essentially fully transmitted electron. 
(Indeed, the triplet case appears to be almost transparent contrary to the singlet case for the higher momentum case considered in Ref.~\cite{SLWM17}).
One interpretation of this result is that the temporary trapping of the incoming electron is
lessened when there is a same-spin electron in the well, as Pauli exchange prevents the
two electrons from being too close, so the incoming electron is not as much disturbed by the atom.

\section{Elastic Scattering via Linear Response TDDFT}
\label{sec:lrtddft}
Just as scattering can be treated in time-independent quantum mechanics, scattering amplitudes can also be extracted from the linear response formalism in TDDFT~\cite{WMB05,FB09,FWEZ07}. Refs.~\cite{FWEZ07,FB09} derived an elegant way to extract elastic scattering cross-sections from calculations of excitations that place the system in a large but {\it finite}-sized infinite-walled box, enabling the straightforward use of standard TDDFT codes. The key realization is that eigenfunctions of a system placed in a box are identical with eigenfunctions of the free system in the interior of the box, for energies that respect the box boundary conditions. Outside the box the eigenfunctions are zero, while in the free system they continue as plane waves, or Coulomb-modified plane waves if the system has a $-1/r$ tail. A given box-size $R$ then filters the continuum of solutions of the free system, supporting only those that go to zero at $R$. 
In Refs.~\cite{FWEZ07,FB09}, the phase-shift, which is the central object in the scattering theory, is extracted in three dimensions from the asymptotic behavior of the wavefunction, using the fact that the wavefunction goes to zero at $R$. 

For one-dimensional scattering, Ref.~\cite{FB09} showed how to generalize their three-dimensional approach via an effective phase-shift analysis in one-dimension presented in Ref.~\cite{E65}. 
Here, instead, we derive an expression for the transmission and reflection coefficients in one-dimensional scattering problems that is inspired by the work of Ref.~\cite{FWEZ07,FB09} but that by-passes the phase-shift and works directly with the coefficients. 
We present the method here, and then show how to use TDDFT in this context, before calculating the transmission and reflection coefficients for our model e-He$^+$ problem. 

Consider one-electron eigenfunctions of a potential that asymptotically goes to zero. For energies above zero, these are doubly-degenerate, and can be written in the form of scattering solutions:
\ben
\psi_{L}(x) \to \left\{ 
\begin{array}{l l}
e^{i(kx - Z \ln (2k|x|)/k)} + re^{-i(kx -Z \ln (2k|x|)/k )} & x<0\\
t e^{i(kx + Z \ln(2k\vert x\vert)/k)} & x > 0 
\end{array}\right.
\een
and 
\ben
\psi_{R}(x) \to \left\{ 
\begin{array}{l l}
t e^{-i(kx - Z \ln(2k\vert x\vert)/k)} & x < 0 \\
e^{-i(kx + Z \ln (2k|x|)/k)} + re^{i(kx +Z \ln (2k|x|)/k )} & x<0
\end{array}\right.
\een
for $\vert x\vert \to \infty$, which comes from solving the time-independent 1D Schr\"odinger equation asymptotically far from the potential, $v\ext \to -Z/|x|$. For scattering off a neutral atom such as in e-H scattering, we take $Z =0$, while for e-He$^+$ scattering, we take $Z = 1$. 
Here $k = \sqrt{2\epsilon}$, with $\epsilon$ the energy eigenvalue.
A general energy eigenfunction can be expressed as a linear combination 
\ben
\psi_{\rm gen} = \mathcal{N}(\psi_L + c \psi_R)
\een
 where $\mathcal{N}$ is a normalization constant and $c$ is a complex constant. 
Now, as in Refs.~\cite{FB09,FWEZ07}, we place the system in a box with infinite walls at $R_-$ and $R_+$, and realize that positive energy solutions to the continuum problem overlap with those in the box in the interior of the box, but only those with that have a node at $R_-$ and $R_+$ can be supported. To simplify, we consider only symmetric potentials, so that solutions in the box have a definite parity, and we choose $R_- = R_+$. For even solutions, it is straightforward to show that $c = +1$, while $c = -1$ for odd solutions. 
Then, requiring that $\psi_{\rm gen}(R_+) = 0$ gives equations that relate $t$ and $r$ for even and odd solutions:
\ben
t \pm r = \mp e^{-2i(kR_{e/o} + Z \ln(2k R_{e/o})/k)} 
\een
where $R_e$ is the radius of the box that supports an even solution of energy $\epsilon = k^2/2$ and $R_o$ is that for an odd solution of this energy. Putting these together, we obtain 
\bea
\nonumber
t = \frac{1}{2}\left(e^{-2i(k R_o + Z \ln(2kR_o)/k)} - e^{-2i(k R_e + Z \ln(2kR_e)/k)}\right)\\
\nonumber
r = -\frac{1}{2}\left(e^{-2i(k R_o + Z \ln(2kR_o)/k)} + e^{-2i(k R_e + Z \ln(2kR_e)/k)}\right)\\
\label{eq:LRtandr}
\eea
It is straightforward to check that $\vert t\vert ^2  + \vert r\vert ^2 = 1$. 

Note that there are an infinite number of box sizes that support an even (or odd) state of momentum $k = \sqrt{2\epsilon}$, and any of these can be used in the formulae Eq.~(\ref{eq:LRtandr}). For any pair $(R_o,R_e)$, the resulting $t$ and $r$ must be the same. This yields identities. For example, consider a fixed $R_o$ and two different $R_e$, where all three boxes support a certain energy eigenstate $\epsilon$. Let $R_e$ and $R_e +\Delta$ be the radii of the two boxes in which this eigenstate is even. Then, for this $k$, $\exp\left(-2i (k\Delta + Z\ln(\frac{R_e+\Delta}{R_e})/k)\right) = 1$. For $\Delta/R_e$ small, this means that
\ben
\Delta = j\frac{\pi}{k + Z/(k R_e)}
\een
where $j$ is an integer. This is a useful relation to check the numerical calculations. 

Eqs.~(\ref{eq:LRtandr}) enable us to extract elastic 1D transmission and reflection amplitudes of potentials with a Coulomb tail from finite-box calculations of eigenstates, similarly to Refs.~\cite{FWEZ07,FB09}. To find these amplitudes for a given momentum, one needs to search for a box radius $R_e$ which supports an even state of energy $\epsilon = k^2/2$ and a box radius $R_o$ that supports an odd state of this energy, and then plug into Eqs.~(\ref{eq:LRtandr}). 

Now we use Eq.~(\ref{eq:LRtandr}) to calculate the reflection and transmission probabilities using AEXX. It was argued in Ref.~\cite{FB09,FWEZ07}, that approximate TDDFT used in this linear response way yields very good scattering cross-sections, provided the approximation has the correct asymptotic behavior in the potential, which AEXX does. Scattering of one electron off an $N$-electron target is computed from excitation energies of the $N+1$ electron system; for our e-H scattering of Ref.~\cite{SLWM17}, this means excitation energies of the 1D H$^-$ ion, and for our e-He$^+$ case, energies of the 1D Helium atom. To get these within TDDFT, we use the matrix equations as derived in the linear response formalism~\cite{C95,C96,PGG96,GPG00} and coded in octopus~\cite{octopus,octopus2}.  The AEXX ground-state of the 1D Helium atom model is first computed in a chosen box radius, and many unoccupied KS orbitals and their orbital energies are computed. These orbital energies are then corrected towards the TDDFT ones, using the xc kernel via the TDDFT linear response matrix equations. The procedure yields excitation frequencies, from which an equivalent incoming momentum is extracted in the following way:
\ben
k = \sqrt{2(\omega + (E_{\rm gs}^{\rm 1D He} - E_{\rm gs}^{\rm 1D He^+}))}
\een
that follows under the assumption that the excitation energy is equal to the ground-state energy of the one-electron target plus the energy of the incoming electron. The parity of this state is then examined. Then the procedure is repeated at a different box radius, for which this excitation frequency corresponds to a state of the opposite parity. 

\begin{figure}[h]
 \centering
 \includegraphics*[width=1.0\columnwidth]{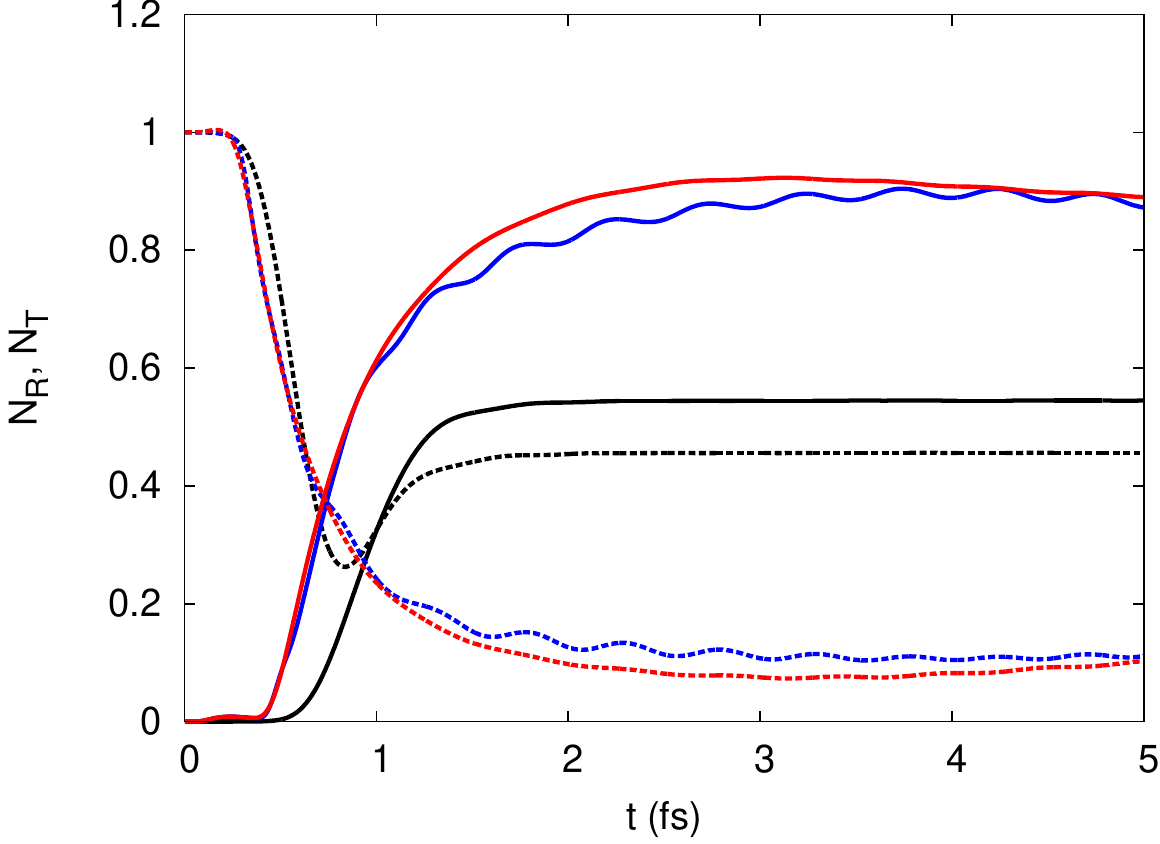}
 \caption{The time-resolved $N_T$ (solid) and $N_R$ (dashed) for the exact (black) and AEXX propagations in the e-H scattering problem with $p=0.6$ and $\Phi_0 = \Phi^{(1)}_0$ (blue)  and $\Phi_0 =\Phi^{(2)}$ (red). 
   Here we choose $\alpha = 0.02$. 
 }
 \label{fig:eH_aexx_exact}
\end{figure}

Following this procedure, we find for the e-H case, for incoming momentum $k = 0.609$a.u., (which is below half the energy of the lowest excitation of the H atom, so well within the elastic scattering regime),  AEXX gives  $ |t|^2 = 0.506$, and $|r|^2 = 0.494$. Comparing with Figure~\ref{fig:eH_aexx_exact}, we see that these numbers are quite close to the {\it exact} probabilities, consistent with the claims of Ref.~\cite{FWEZ07,FB09}, while very different from results propagated using AEXX which are closer to 0.9 and 0.1, respectively, when using either  $\Phi^{(1)}$ or $\Phi^{(2)}$.  That is, the AEXX used in the full real-time propagation calculation gives a completely different result than AEXX used in the linear response calculation. Note that in this time-resolved calculation, we reduced the wavepacket width $\alpha = 0.02$ instead of the 0.1 we used earlier, in order to reduce the range of momenta that make up the initial wavepacket (i.e. closer to the plane-wave limit).

\begin{figure}[h]
 \centering
 \includegraphics*[width=1.0\columnwidth]{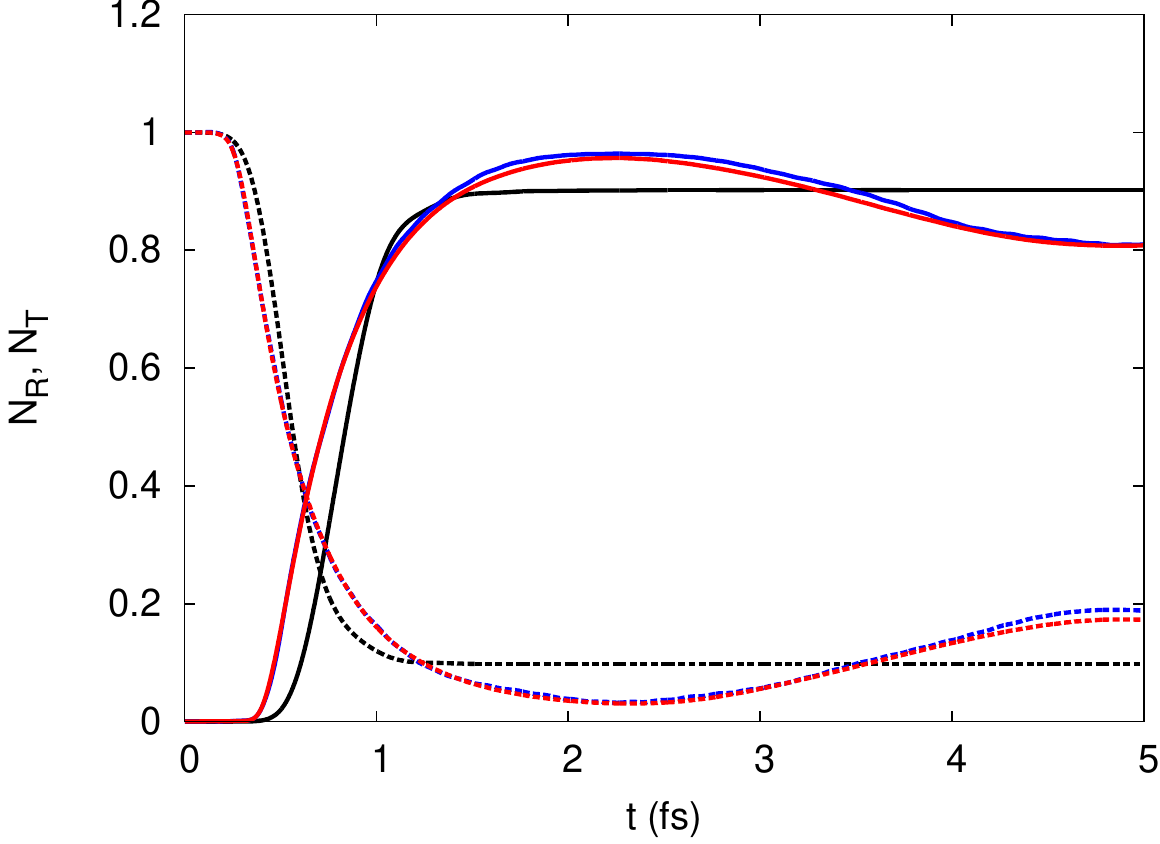}
 \caption{The exact time-resolved $N_T$ (solid) and $N_R$ (dashed) for the exact (black) and AEXX propagations in the e-He$^+$ scattering problem with $p=0.6$ and $\Phi_0 = \Phi^{(1)}_0$ (blue) and $\Phi_0 =\Phi^{(2)}$ (red).    Here we choose $\alpha = 0.02$.}
 \label{fig:eHep_aexx_exact}
\end{figure}

In the case of e-He$^+$ scattering, for an incoming momentum $k = 0.6085$a.u., AEXX gave $ |t|^2 = 0.89$, and $|r|^2 = 0.11$. Comparing this to our wavepacket calculation, again adjusting the width to $\alpha = 0.02$a.u., we plot the results of time-propagation using AEXX in Figure~\ref{fig:eHep_aexx_exact}.  AEXX used within linear response is again very close to the exact value of $N_T$ and $N_R$, consistent with the claim of Refs.~\cite{FWEZ07,FB09} that scattering probabilities can be well-approximated from adiabatic TDDFT. In this particular case, the integrated quantities of $N_R$ and $N_T$  in AEXX in full time-propagation of the doubly-occupied state $\Phi^{(1)}$ are not that bad but not that good either (see discussion Sec.~\ref{sec:approx}).

As we have seen in the previous section, and in Ref.~\cite{SLWM17}, AEXX does generally quite badly in predicting the scattering dynamics in real time. Here we have explicitly verified the claim we made in Ref.~\cite{SLWM17}: an adiabatic functional may be able to predict reasonably good elastic scattering amplitudes from a linear response calculation but it will tend to underestimate scattering when used in a fully time-resolved way. 
This is because, as argued in Ref.~\cite{SLWM17}, the two situations probe different regions of the functional: in the former, the system is merely perturbed away from the ground-state, while in the latter we leave the ground-state from the very beginning of the dynamics. 
The exact  time-resolved xc potential shows distinctive peak and valley structures in both the elastic and inelastic cases, absent in AEXX.
Further,  when scattering is studied in the linear-response way, only elastic scattering can be extracted. (In fact, if we tried to apply the procedure above to a higher $k$ such as $k = 1.2$ we get $N_T = 0.99995$, almost perfect transmission, while the time-resolved dynamics clearly shows appreciable reflection).

\section{Summary and Outlook}
\label{sec:outlook}
Both models of electron-atom and electron-ion scattering have shown similar features in the exact xc potential: peaks and valley structures that are necessary to obtain a qualitatively correct time-evolution as well as good reflection and transmission coefficients. Approximations that lack these structures  tend to underestimate reflection, typically quite significantly, and one must go beyond the adiabatic approximation in order to capture them~\cite{SLWM17}. Our results are based on 1D models so cannot capture effects from channels where the electron scatters around the target in real systems. Also, whether the errors from the approximate methods are so large for realistic systems with more electrons and vibronic effects, is unclear; still, the results here suggest the tendency of adiabatic TDDFT to underestimate scattering in realistic systems~\cite{GWWZ14,QSAC17,Kirchner,scat3,scat4,scat5,scat6}.

The choice of Kohn-Sham initial state can greatly affect the accuracy of the dynamics predicted by a TDDFT approximation. Unphysical oscillations in the density can be avoided if one chooses an initial KS state 
"close" to the physical state. When instead a Slater determinant with a doubly-occupied orbital is chosen to represent the scattering of an electron from a target, to compensate the resulting unphysical interferences, the KS potential will have to create complex structures to spatially split the wave-function and these structures are impossible to reproduce for any actual approximation of the potential. 

If the initial state is chosen well, using the relatively new non-adiabatic approximation $v\xc^{\rm S}$ propagates very well up until the collision really starts but it ultimately fails as it ignores a large part of the correlation. The missing part, $v\c^T$, contains the elaborate shapes of the potential, and an exact expression is known for it, in terms of the difference between the exact and Kohn-Sham one-body reduced density matrices. To our knowledge, no convincing approximation of this term has been proposed, as a functional of the initial states and density, although work in this direction is underway. 

In the elastic regime scattering probabilities can also be extracted from TDDFT linear response, for which the conventional functionals fare much better. We have derived an expression for the transmission and reflection coefficients for the 1D scattering scenario, along the lines of the ideas of Refs~\cite{FWEZ07,FB09}. We explicitly showed that, used in this context, AEXX works quite well, even though AEXX in the fully time-resolved picture fares much worse, and have argued that this is because the latter involves evaluating the functional very far from the ground-state. To get truly reliable and accurate scattering cross-sections, one must go beyond the adiabatic approximation.

\begin{acknowledgement}
YS is supported by JSPS KAKENHI Grant No. JP16K17768.
KW is supported by JSPS KAKENHI Grant No. JP16K05483.
Financial support from the US National Science Foundation
CHE-1566197 (NTM) and the Department of Energy, Office
of Basic Energy Sciences, Division of Chemical Sciences,
Geosciences and Biosciences under Award DE-SC0015344 (LL)
are also gratefully acknowledged.
Part of the computations were performed on
the supercomputers of the Institute for Solid State Physics,
The University of Tokyo.
\end{acknowledgement}

\bibliographystyle{epj}
% \bibliography{}
\bibliography{./scattering}

\end{document}